\documentclass[journal=jpcbfk,manuscript=article]{achemso}

\usepackage{natmove}

\author{Caroline Desgranges}
\author{Leanna Widhalm}
\author{Jerome Delhommelle}
\email{jerome.delhommelle@und.edu}
\affiliation[University of North Dakota]
{Department of Chemistry, University of North Dakota, Grand Forks, ND 58202, USA\\
(Phone: 701-777-2495)}

\title{Scaling Laws and Critical Properties for $FCC$ and $HCP$ Metals.}

\begin{document}

\begin{abstract}
The determination of the critical parameters of metals has remained particularly challenging both experimentally, because of the very large temperatures involved, and theoretically, because of the many-body interactions that take place in metals. Moreover, experiments have shown that these systems exhibit an unusually strong asymmetry of their binodal. Recent theoretical work has led to new similarity laws, based on the calculation of the Zeno line and of the underlying Boyle parameters, which provided results for the critical properties of atomic and molecular systems in excellent agreement with experiments. Using the recently developed Expanded Wang-Landau (EWL) simulation method, we evaluate the grand-canonical partition function, over a wide range of conditions, for $11$ $FCC$ and $HCP$ metals ($Ag$, $Al$, $Au$, $Be$, $Cu$, $Ir$, $Ni$, $Pb$, $Pd$, $Pt$ and $Rh$), modeled with a many-body interaction potential. This allows us to calculate the binodal, Zeno line, Boyle parameters and, in turn, obtain the critical properties for these systems. We also propose two scaling laws for the enthalpy and entropy of vaporization, and identify critical exponents of $0.4$ and $1.22$ for these two laws, respectively.
\end{abstract}

\section{Introduction}
The determination of the critical properties often relies on the extrapolation~\cite{Rowlinson,panagiotopoulos1987direct,bhatt2006critical,FitCrit,Potoff,Errington,messerly2015improved,Vrabec} of low temperature results using e.g., for the critical density, the law of rectilinear diameter~\cite{Cailletet}. This law is based on the observation of the linearity of the diameter $\rho_m$ of the liquid-vapor coexistence curve as a function of temperature, with $\rho_m=0.5(\rho_l+\rho_v)$ where $\rho_l$ and $\rho_v$ are the liquid density and vapor density at coexistence, respectively. Although this law seems to hold for a wide range of substances~\cite{widom1970new,PhysRevB.42.6651}, there have been exceptions~\cite{PhysRevLett.32.879,PhysRevLett.85.696,sengers2009experimental} most notably in the case of metals~\cite{Jungst,Goldstein,schroer2014estimation}. For instance, for alkali metals like $Cs$ and $Rb$, experiments showed that the two branches of the coexistence curves were strongly asymmetric and the law of rectilinear diameter was found to break down over a large temperature range~\cite{Jungst}. The exceptional nature of metals was attributed to the existence of many-body effects in these systems~\cite{Goldstein}, which have been shown to play a major role close to the critical region~\cite{Goldstein2}.

The knowledge of the critical properties of metals is of key importance for many applications at the nanoscale~\cite{bhatt2006critical,bhatt2006phase} and  for high temperature technologies~\cite{schroer2014estimation}, e. g. in aerospace and in the nuclear industry. However, there is generally a lack of experimental data on metallic systems. This is due to the large temperatures involved in the determination of the critical parameters of metals (the liquid range of metals can cover temperatures which may go above $10000~K$~\cite{schroer2014estimation}). These are particularly challenging to investigate and require special experimental techniques like the "exploding-wire" technique~\cite{pottlacher1996review,pottlacher1993investigations}. As a result, the extrapolations from the experimental data exhibit large variations~\cite{schroer2014estimation,lang,chhabra1990surface,kaptay}. For instance, estimates for the critical temperature of $Al$ vary from around $T_c=5500~K$ to $T_c=9600~K$~\cite{morel2009critical}. Similarly for $Cu$, the estimated critical temperatures range from $5100~K$ to $8900~K$~\cite{ramana2014molecular,singh2006vapor}. To bridge this gap in knowledge, recent work has led to the determination of the critical properties from low temperature liquid data. These studies either used a power series law for the diameter~\cite{schroer2014estimation} or a new symmetrized equation for the vapor liquid coexistence curve~\cite{apfelbaum2015wide,apfelbaum2012estimate,apfelbaum2014saturation,apfelbaum2008new,apfelbaum2006triangle,kulinskii2010simple,kulinskii2014critical,Vorobev}.

The aim of this work is to use molecular simulation to determine the critical properties of a series of $FCC$ and $HCP$ metals modeled with a many-body force field known as the quantum-corrected Sutton-Chen potential~\cite{luo2003maximum}. This model was shown to accurately model the properties of liquid metals~\cite{luo2003maximum,kart2005thermodynamical,xu2005assessment, qi2001viscosities,kart2004liquid,kart2006structural,desgranges2007viscosity,desgranges2008shear,desgranges2008rheology}, as well as the boiling points of metals~\cite{gelb2011boiling} and the vapor-liquid equilibrium properties of $Cu$~\cite{Tsvetan,Tsvetan2}. To determine the vapor-liquid properties, as well as the locus of the Zeno line, we use the recently developed Expanded Wang-Landau simulation method~\cite{PartI,PartII,PartIII,PartIV}. This method leads to an accurate determination of the grand-canonical partition function of a system, and in turn, to all thermodynamic properties, including the vapor and liquid densities at coexistence and the compressibility factor. This allows us to determine the Zeno line, the Boyle parameters and the critical point for all metallic systems studied in this work.

The paper is organized as follows. In the next section, we present the simulation methods as well as the many-body force field used in this work. We show the results obtained for $11$ $FCC$ and $HCP$ metals and explain how we determine the critical properties of these systems. We then discuss and compare our results to those obtained by extrapolating experimental data at low temperature. We finally draw the main conclusions from this work in the last section.

\section{Simulation methods}

The fluid properties at coexistence are determined using the recently developed Expanded Wang-Landau simulations (EWL)~\cite{PartI,PartII,PartIII,PartIV}. We give here an outline of the method. The EWL simulation method relies on a combination of a Wang-Landau sampling scheme~\cite{Wang1,Wang2,Shell,Yan,Camp,WLHMC,Tsvetan,DHMD,KennethI,KennethII,Tsvetan2} with an expanded ensemble approach~\cite{expanded,Lyubartsev,Paul,Shi,Singh,MV1,MV2,Rane1,Rane2,Aaron,Erica,Andrew,jctc2015} to provide an accurate value of the grand-canonical partition function for a given system. The use of a Wang-Landau sampling allows to determine iteratively the biasing function (hence the partition function), while the use of an expanded ensemble approach, which consists in dividing the insertion/deletion of an atom into $M$ stages, ensures that the grand-canonical ensemble is accurately sampled over the whole range of conditions. This is especially key for high density liquids~\cite{PartI} like liquid metals. The output from the simulation are the grand-canonical partition function $\Theta(\mu,V,T)$ and the $Q(N,V,T)$ function which are related as follows:
\begin{equation}
\Theta(\mu,V,T)= \sum_{N=0}^\infty Q(N,V,T) \exp (\beta \mu N)\\
\label{muVTfinal}   
\end{equation}
with $Q(N,V,T)$ defined as 
\begin{equation}
Q(N,V,T)= {V^{N} \over { N! \Lambda^{3N}}}  \int \exp\left(-\beta U({ {\Gamma}})\right) d{ {\Gamma}} \\
\label{Q_NVTfrac}   
\end{equation}
where $\mu$ denotes the chemical potential, $V$ the volume, $T$ the temperature, $N$ the number of atoms in the system, $\Lambda$ the De Broglie wavelength, $U$ is the potential energy of the system at a given point $\Gamma$ of the configuration space.

Once the partition function is known, all thermodynamic properties of the system can be calculated through the statistical mechanics formalism~\cite{McQuarrie} and the number distribution $p(N)$ can be calculated as:

\begin{equation}
p(N) = {Q(N,V,T)  \exp\left(\beta \mu N\right) \over \Theta(\mu,V,T)}\\
\label{pN}   
\end{equation}

This allows to determine the densities at coexistence as well as the other thermodynamic properties at coexistence~\cite{PartI,PartII,PartIII,PartIV}, including pressure $P$, enthalpies $H_{liq}$ and $H_{vap}$ as well as entropies $S_{liq}$ and $S_{vap}$. From there, the location of the Zeno line, i.e. the points for which the compressibility factor $Z=PV/RT=1$ can be readily determined.

The many-body interactions in metals are often taken into account using embedded-atom potentials (EAM)~\cite{finnis1984simple,sutton1990long,mei1991analytic,daw1983semiempirical}. EAM potentials are density-dependent potentials that were initially introduced to model the properties of the solid phases of metals. In this work, we use the quantum-corrected Sutton-Chen~\cite{luo2003maximum} embedded atoms (qSC-EAM) potential. The qSC-EAM potential has been shown to accurately describe thermodynamic properties of liquid metals~\cite{luo2003maximum,kart2005thermodynamical,xu2005assessment, qi2001viscosities,kart2004liquid,kart2006structural,desgranges2007viscosity,desgranges2008shear,desgranges2008rheology}, as well as the boiling points of metals~\cite{gelb2011boiling} and the vapor-liquid equilibrium properties of Copper~\cite{Tsvetan,Tsvetan2}. In the qSC-EAM potential, the potential energy $U$ of a system containing $N$ atoms is written as the sum of a contribution of two-body term and a contribution of a many-body term

\begin{equation}
U={1 \over 2} \sum_{i=1}^N \sum_{j \ne i}{\varepsilon \left( {a \over r_{ij}} \right)^n} -\varepsilon C \sum_{i=1}^N \sqrt \rho_i
\label{total}   
\end{equation}

in which $r_{ij}$ is the distance between two atoms $i$ and $j$ and the density term $\rho_i$ is given by

\begin{equation}
\rho_i=\sum_{j \ne i}{\left( {a \over r_{ij}} \right)^m}
\label{dens}   
\end{equation}

\begin{figure}
\begin{center}
\includegraphics*[width=7.5cm]{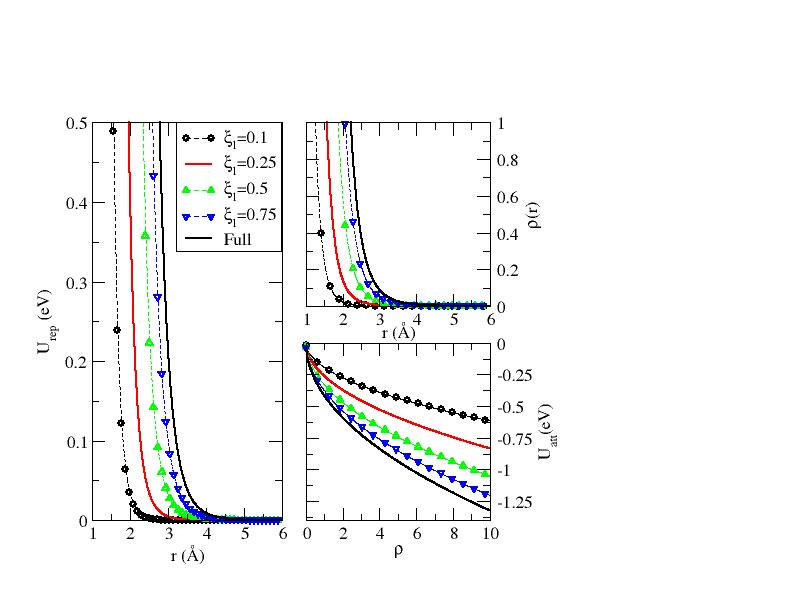}
\end{center}
\caption{Interaction between a fractional atom and a full atom. Results are shown for different values of $\xi_l=l/M$ for the 2-body contribution (left), the local density (top right) and the many-body contribution (bottom right).}
\label{Fig1}
\end{figure}

We use  the parameters of Luo {\it et al.}~\cite{luo2003maximum} for the $11$ $FCC$ and $HCP$ metals ($Ag$, $Al$, $Au$, $Be$, $Cu$, $Ir$, $Ni$, $Pb$, $Pd$, $Pt$ and $Rh$) studied in this work. The interaction between a fractional atom and a full atom is given by Eqs.~\ref{total} and \ref{dens} using the scaled paramters $a_l=a(l/M)^{1/4}$ instead of $a$ and $C_l=C(l/M)^{1/3}$ instead of $C$ where $l$ is the current stage value of the fractional particle $(0<l<M-1)$. Fig.~\ref{Fig1} shows the dependence of the potential for the full-fractional interaction as a function of the stage value for the fractional particle. From a practical standpoint, EWL simulations are carried out within the framework of Monte Carlo (MC) simulations. To sample the configurations of the system, we perform MC steps corresponding to the translation of a single atom ($75\%$ of the MC steps) or to a change in $(N,l)$ for the system ($25\%$ of the MC steps). The EWL method yields accurate results, with error bars~\cite{PartI} typically of the order of 0.1\% for the density and of 0.02~\% for the enthalpy and entropy. The technical details regarding the Wang-Landau scheme are exactly the same as previously described~\cite{PartI,PartII}, with the final value of the convergence factor set to $f=10^{-8}$ and a number of intermediate stages set to $M=100$. 

\section{Results and discussion}
\begin{figure}
\begin{center}
\includegraphics*[width=7.5cm]{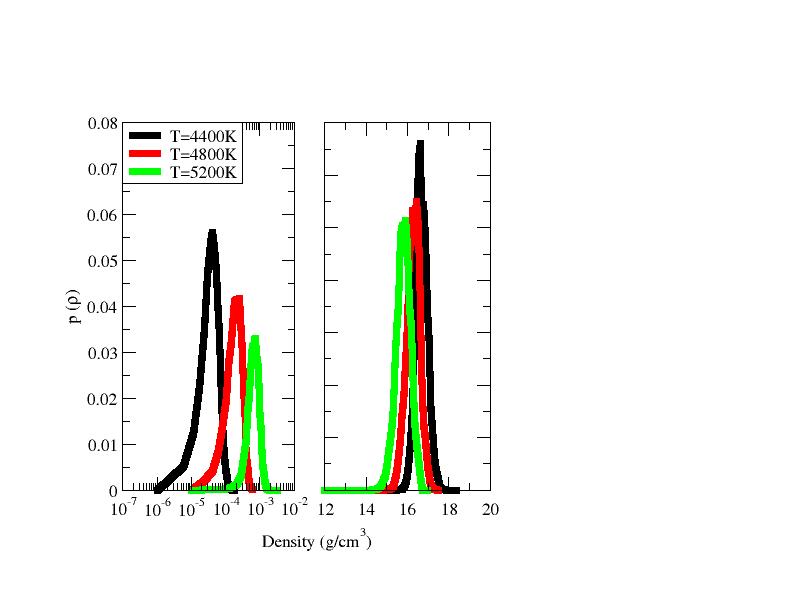}
\end{center}
\caption{$Ir$: density distribution $p(\rho)$ for the vapor (left) and for the liquid (right) along the coexistence line.}
\label{Fig2}
\end{figure}

We start by presenting, on the example of $Ir$, the results obtained using the EWL approach. Fig.~\ref{Fig2} shows the density distributions $p (\rho=N/V)$ obtained at coexistence for three temperatures, $T=4400~K$, $T=4800~K$ and $T=5200~K$. In the left panel of Fig.~\ref{Fig2}, we plot the density distributions for the vapor phase while the right panel of Fig.~\ref{Fig2} shows the density distribution for the liquid phase. These plots exhibit the expected behavior for the distributions, i.e. a shift towards the higher densities for the vapor peak as temperature increases and a shift towards the lower densities for the liquid peak as $T$ increases. 

\begin{figure}
\begin{center}
\includegraphics*[width=7.5cm]{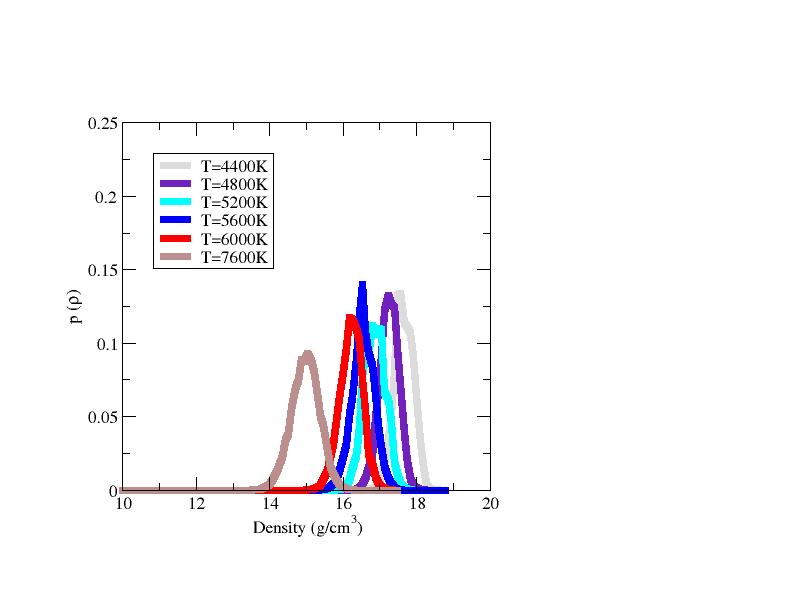}
\end{center}
\caption{$Ir$: density distribution $p(\rho)$ along the Zeno line.}
\label{Fig3}
\end{figure}

We now turn to the density distributions obtained for $Ir$ along the Zeno line. We recall that this is obtained, for each temperature, by identifying the value of the chemical potential that ensures that the ratio $Z=PV/RT$ is equal to $1$. The density distribution corresponding to the chemical potential so obtained is plotted, for each temperature, in Fig.~\ref{Fig3}. We see that in this case we obtain a single peak (corresponding to a single phase system) which gets shifted towards the lower densities as $T$ increases. 

\begin{figure}
\begin{center}
\includegraphics*[width=7.5cm]{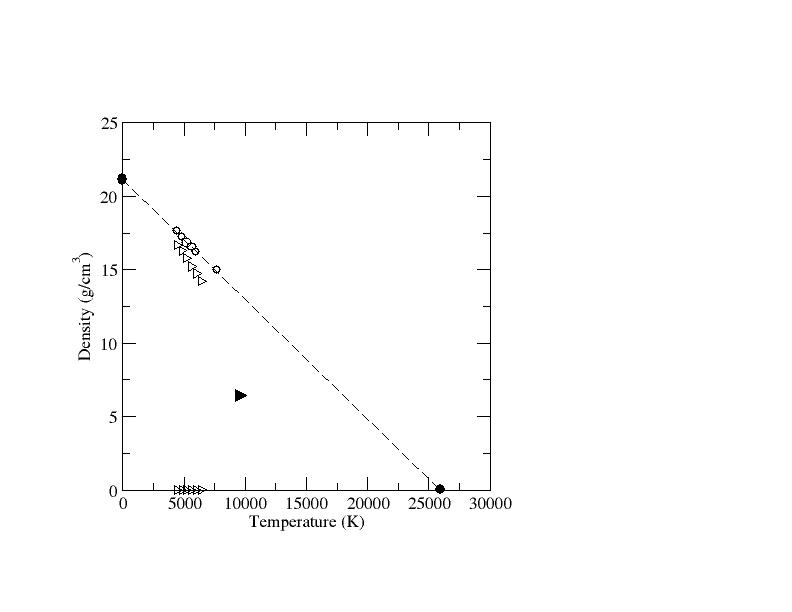}
\end{center}
\caption{$Ir$: EWL results for the vapor-liquid equilibrium curve (open triangles) and for the Zeno line (open circles). The critical point is shown as a filled triangle while the Boyle parameters are shown with filled circles.}
\label{Fig4}
\end{figure}

The phase diagram in the $T-\rho$ plane, as well as the location of the Zeno line in that plane can be readily determined from the density distributions. The complete results for $Ir$ are indicated in Fig.~\ref{Fig4} with the densities at coexistence and the locus of the Zeno line. The results confirm that the Zeno line is straight over the range of liquid densities considered in this work~\cite{fokin2013general}. This plot also shows the Boyle parameters for $Ir$ as well as the critical point. The Boyle parameters are determined by performing a linear fit to the EWL results along the Zeno line. The critical parameters are then determined from the EWL results as follows. First, the critical temperature is extrapolated from the densities of the two coexisting phases through the following fit:

\begin{equation}
\rho_l-\rho_v=A (T_c-T)^{0.326}
\label{ExTc}   
\end{equation}

where $A$ and $T_c$ are two fitting parameters, $\rho_l$ and $\rho_v$ are the densities for the liquid and vapor phases at a given temperature $T$ and the $3$D-Ising critical exponent, adjusted for real substances, of $0.326$ is used.

Second, the critical density $\rho_c$ is obtained using the following similarly relation~\cite{apfelbaum2012estimate}:
\begin{equation}
{{T_c} \over {T_B}} + {{\rho_c} \over {\rho_B}} = S_1
\label{Exrc}   
\end{equation}
where $T_c$ is the critical temperature determined using the previous equation, $T_B$ and $\rho_B$ are the Boyle temperature and density, $S_1$ is a parameter equal to $0.67$ (as established by Apfelbaum {\it et al.} for several systems~\cite{apfelbaum2009predictions}, including $Be$, $Cu$ and $Al$), and $\rho_c$ is a fitting parameter. Using this similarity law alleviates the need to use the law of rectilinear diameter, that has been shown to break down for metals~\cite{Jungst}, to determine the critical density of metals

We apply the same approach to all $11$ $FCC$ and $HCP$ metals considered in this work. Table~\ref{Boyle} summarizes the results obtained in this work both for the Boyle and critical parameters. In addition to the critical temperatures and densities, we also provide an estimate for the critical pressure. The critical pressure is obtained by fitting an Antoine law to the results obtained for the pressure. The fit to the Antoine law is carried out according to:

\begin{equation}
log_{10}~P=A+{B \over {T+C}}
\label{ExPc}   
\end{equation}

where $A$, $B$ and $C$ are fitting parameters and $P$ is the vapor pressure for a given temperature $T$. The critical pressure is calculated from this law using the value for the critical temperature previously obtained from Eq.~\ref{ExTc}.

\begin{table}[hbpt]
\caption{Boyle and critical parameters obtained in this work for $FCC$ and $HCP$ metals.}
\begin{tabular}{|c|c|c|c|c|c|}
\hline
\hline
$$& $T_{B}$&  $\rho_{B}$ & $T_c$ & $\rho_c$ &  $P_c$  \\
$ $&  $(K)$ & $(g/cm^3)$ & $(K)$ &  $(g/cm^3)$ & $(MPa)$  \\
\hline
\hline
$Ag$ & $11488$ &  $9.97$ & $4260$ & $2.98$ & $34.3$  \\
$Al$ & $15316$ &  $2.56$ & $5412$ & $0.81$ & $37.9$  \\
$Au$ & $12627$ &  $18.02$ & $4286$ & $5.96$ & $18.6$  \\
$Be$ & $12534$ &  $1.71$ & $4623$ & $0.52$ & $88.6$  \\
$Cu$ & $15688$ &  $8.36$ & $5430$ & $3.33$ & $57.8$  \\
$Ir$ & $25974$ &  $21.15$ & $9484$ & $6.45$ & $91.7$  \\
$Ni$ & $19502$ &  $8.35$ & $6700$ & $2.72$ & $62.1$  \\
$Pb$ & $8287$ &  $10.16$ & $2663$ & $3.54$ & $7.2$  \\
$Pd$ & $14888$ &  $11.29$ & $5444$ & $3.44$ & $50.1$  \\
$Pt$ & $21472$ &  $20.00$ & $7375$ & $6.53$ & $44.5$  \\
$Rh$ & $21691$ &  $11.72$ & $8106$ & $3.47$ & $62.1$  \\
\hline
\hline
\end{tabular}
\label{Boyle}
\end{table}

We first discuss the results obtained for $Be$. We start by comparing the EWL results for the Boyle parameters to those obtained in previous work by Apfelbaum~\cite{apfelbaum2012estimate} using an effective ion-ion potential to model liquid $Be$. We obtain a very good agreement for the Boyle density between the EWL results on the qSC-EAM potential ($1.714~g/cm^3$) and the previous estimate ($1.697~g/cm^3$), and a reasonable agreement for the Boyle temperature with a EWL value of $12534~K$ compared to $10500~K$. The critical temperature (see Table~\ref{Comp}) we estimate from our EWL results ($T_c=4623~K$) is in reasonable agreement with the predicted value of $5400~K$ by Apfelbaum~\cite{apfelbaum2012estimate}, obtained by analyzing the low temperature liquid data for $Be$. Using the similarity law based on the Zeno line, we find a critical density of $0.516~g/cm^3$. From the fit to an Antoine law, we obtain the following estimate for the critical pressure $P_c=886~bar$, which is comparable to the results obtained in previous work~\cite{apfelbaum2012estimate}.

Focusing now on the case of $Al$, we obtain results for the Boyle parameters that share the same features as for $Be$. The Boyle density we obtain from the EWL simulations ($2.56~g/cm^3$) is once again in very good agreement with the results from Apfelbaum and Vorob'ev~\cite{apfelbaum2009predictions} ($2.57~g/cm^3$), while we obtain a larger Boyle temperature (with a EWL value of $15316~K$ compared to $12888~K$). For the critical parameters, we have a critical temperature estimated at $T_c=5412~K$, which is approximately a thousand $K$ below the previous estimate $T_c=6318~K$, made on the basis of the liquid data, and a critical density that is larger than the previous estimate ($0.45~g/cm^3$ versus $0.81~g/cm^3$ in this work). The deviation for the critical density can be accounted for the use of a lower value for the critical temperature in the similarity law (Eq.~\ref{Exrc}). More specifically, the EWL results give a $T_c:T_B$ ratio of $0.35$, below the ratio of $0.49$ found in previous work~\cite{apfelbaum2009predictions}, resulting in a greater value for the $\rho_c:\rho_B$ ratio as fixed by the similarity law. Comparing now our results to those obtained using Gibbs Ensemble Monte Carlo simulations on a different EAM potential~\cite{bhatt2006critical} and the law of rectilinear diameter, we find that our estimate for the critical density is much closer to that from previous simulation work ($0.707~g/cm^3$) and that our estimate for the critical temperature is  $14\%$ below those results.

In the case of $Cu$, the Boyle parameters we obtain from the EWL simulations on the qSC-EAM potential are in very good agreement with the findings from previous work~\cite{apfelbaum2009predictions} both for the Boyle density $(8.6~g/cm^3$ compared to $8.36~g/cm^3$ in this work) and for the Boyle temperature ($15593~K$ compared to $15688~K$ in this work). In line with the results for $Al$, the EWL critical temperature ($T_c=5430~K$) is below the value extrapolated from the liquid data ($T_c=7093~K$), leading to a larger critical density, since the same similarity law, with different input values for $T_c$, is used. 

We finally compare the critical parameters obtained in this work for $Pb$ and $Au$ to those estimated from the experimental data on liquid densities by Schroer and Pottlacher~\cite{schroer2014estimation}, which rely on averaging the results obtained using either $3D$-Ising scaling law or a mean-field scaling law.  For $Pb$, the critical temperature we estimate is $T_c=2663~K$ (as compared to $4636~K$ in prior work~\cite{schroer2014estimation}), while the critical density we estimate is $\rho_c=3.54~g/cm^3$ is above the critical density from previous work~\cite{schroer2014estimation} ($\rho_c=2.322~g/cm^3$). For $Au$, we estimate the critical temperature to be $T_c=4286~K$ (as compared to $7217~K$ in prior work~\cite{schroer2014estimation}) while the critical density we estimate is $\rho_c=5.96~g/cm^3$ is above the critical density from previous work~\cite{schroer2014estimation} ($\rho_c=3.544~g/cm^3$). The critical temperatures we estimate are therefore below those extrapolated in previous work~\cite{schroer2014estimation}, while the critical densities we estimate are larger. This is due to two notable differences between the scaling laws used for the critical temperature. First, in our case, we are able to obtain EWL results for high temperature phases and get an accurate estimate for the variations of the density order parameter ($\rho_l-\rho_v$), over a wide temperature range while the extrapolation from the low temperature data only takes into account the variation of $\rho_l$ over a limited temperature range. Second, we use, for the density order parameter, a scaling law with an exponent of $0.326$, while the extrapolation from the data~\cite{schroer2014estimation} calculates the average result for two scaling laws applied to the liquid density $\rho_l$, one with a $3D$-Ising exponent and one with a mean-field exponent of $0.5$.

\begin{table}[hbpt]
\caption{Comparison of the critical parameters for selected metals ($^a$: this work, $^b$: Apfelbaum~\cite{apfelbaum2012estimate}, $^c$:  Apfelbaum and Vorob'ev~\cite{apfelbaum2009predictions},  $^d$: Schroer and Pottlacher~\cite{schroer2014estimation}, $^e$: Kaptay~\cite{kaptay} and $^f$: Bhatt {\it et al.} ~\cite{bhatt2006critical}.}
\begin{tabular}{|c|c|c|c|c|c|}
\hline
\hline
$$& $T_c$ & $\rho_c$ &  $P_c$  \\
$$& $(K)$ &  $(g/cm^3)$ & $(MPa)$  \\
\hline
\hline
$Be$ & $4623^a$ & $0.52^a$ & $88.6^a$  \\
 & $5400^b$ & $0.26^b$ & $46.6^b$  \\
 & $10500^e$ & &   \\
\hline
$Al$ & $5412^a$ & $0.81^a$ & $37.9^a$  \\
 & $6378^c$ & $0.45^c$ & $108.8^c$  \\
 & $6299^f$ & $0.71^f$ & $89.5^f$   \\
\hline
$Cu$ & $5430^a$ & $3.33^a$ & $57.8^a$  \\
 & $7093^c$ & $1.95^c$ & $45.6^c$  \\
\hline
$Au$ & $4286^a$ & $5.96^a$ & $18.6^a$  \\
 & $7217^d$ & $3.54^d$ &  \\
 & $8700^e$ & &  \\
\hline
$Pb$ & $2663^a$ & $3.54^a$ & $7.2^a$  \\
 & $4836^d$ & $2.32^d$ &  \\
 & $4200^e$ & &  \\
\hline
\hline
\end{tabular}
\label{Comp}
\end{table}

\begin{figure}
\begin{center}
\includegraphics*[width=7.5cm]{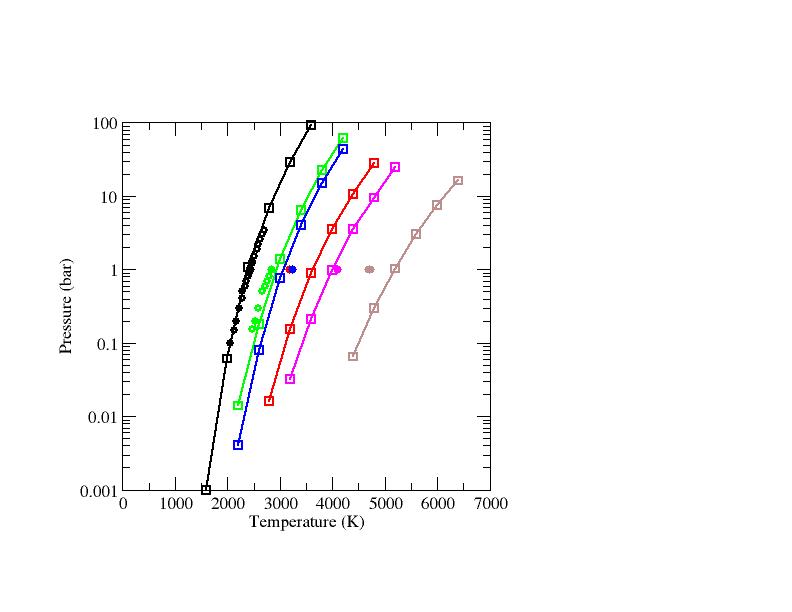}
\end{center}
\caption{Vapor pressure of selected metals. EWL results are shown as open squares ($Ag$ in black, $Cu$ in green, $Pd$ in blue, $Ni$ in red, $Pt$ in magenta and $Ir$ in brown) while experimental data are shown with open circles~\cite{geiger1987vapor} and filled circles~\cite{linde1994handbook}.}
\label{Fig5}
\end{figure}

We provide in Fig.~\ref{Fig5} a comparison between the EWL result for the vapor pressure and the experimental data. These results show that there is generally a good agreement between the simulation results obtained with the qSC-EAM potential as can be seen most notably in the case of $Ag$ and $Cu$. As previously discussed~\cite{gelb2011boiling}, the qSC-EAM potential leads to reliable results for the boiling points in most cases (see e.g. $Ag$, $Cu$ and $Pt$ on the plot), with a few exceptions (e.g. with deviations of up to $10\%$ for $Ni$ and $Ir$ for the boiling temperatures in Fig.~\ref{Fig5}).  

\begin{figure}
\begin{center}
\includegraphics*[width=7.5cm]{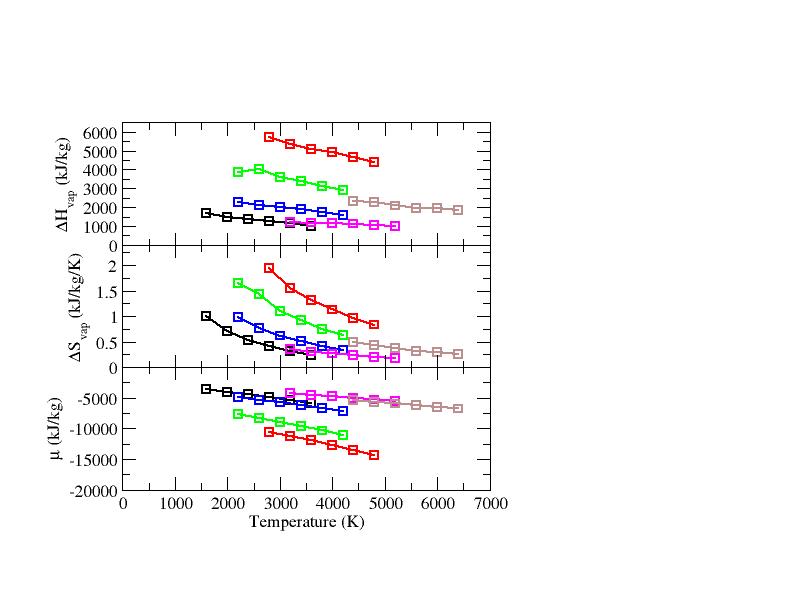}
\end{center}
\caption{Thermodynamic properties at coexistence for selected metals: $\Delta H_{vap}$ against temperature (top), $\Delta S_{vap}$ against $T$ (middle) and $\mu$ against $T$ (bottom). Same legend as in Fig.~\ref{Fig5}.}
\label{Fig6}
\end{figure}

The EWL results for the variations of the other thermodynamic properties at coexistence are plotted against temperature in Fig.~\ref{Fig6}. All metals exhibit the same qualitative behavior with the linear variation of $\mu$ as a function of $T$ (see bottom of Fig.~\ref{Fig6}), and the expected decrease of $\Delta H_{vap}$ (see top of Fig.~\ref{Fig6}) and of $\Delta S_{vap}$ (see middle of Fig.~\ref{Fig6}) towards $0$ as the temperature approaches the critical temperature. To further analyze the results, we fit the simulation results for the enthalpy of vaporization with the following function:

\begin{equation}
\Delta H_{vap}=A (T_c-T)^a
\label{SLH}   
\end{equation}

where $A$ and $a$ are two fitting parameters and $T_c$ is the critical temperature previously determined. Carrying out this fit over the set of metals considered here, we find an average critical exponent for $\Delta H_{vap}$ of $0.4$. The value for this exponent is very close to the value of 0.38 found in prior work on non-metals~\cite{watson1943thermodynamics,leibovici2014new}. We apply the same analysis to the results for $\Delta S_{vap}$ with the following function:

\begin{equation}
\Delta S_{vap}=B (T_c-T)^b
\label{SLS}   
\end{equation}

where $B$ and $b$ are two fitting parameters and $T_c$ is the critical temperature previously determined. The exponent we find in this case is $1.22$. To our knowledge, this provides the first determination, in the case of metals, of the critical exponents for these two scaling laws.

\section{Conclusions}
In this work, we carry out EWL simulations to determine the binodal and the Zeno line of $11$ $FCC$ and $HCP$ metals ($Ag$, $Al$, $Au$, $Be$, $Cu$, $Ir$, $Ni$, $Pb$, $Pd$, $Pt$ and $Rh$), modeled with a many-body potential known as the qSC-EAM model. We estimate the critical parameters according to a three-step procedure with (i) the calculation of the critical temperature from a scaling law for the density (with a critical exponent of $0.326$), (ii) the determination of the critical density using a similarity law and the Boyle parameters defining the Zeno line, and (iii) the evaluation of the critical pressure from an Antoine law fitted to the simulation results for the vapor pressure. We compare our results to estimates made in prior work on the basis of the experimental data for liquid densities and, when available, to previous simulation work that relied on the law of rectilinear diameters. The results obtained with our EWL simulations cover a wider range of temperatures than that covered in the extrapolations from the liquid densities data. They also take into account the variations of the density order parameter ($\rho_l-\rho_v$) at high temperature, which are not included when the extrapolation from low temperature liquid data only is carried out. Applying the three-step procedure to the $11$ $FCC$ and $HCP$ metals studied in this work allows us to provide an estimate for the critical parameters for all systems. These estimates generally lead to lower critical temperatures and, as a result, to larger critical densities than those obtained from extrapolations relying on low temperature liquid data only. Our results also allow us to propose two scaling laws for the enthalpy of vaporization and the entropy of vaporization with the corresponding critical exponents of $0.4$ and $1.22$ respectively. Further work testing the validity of these two scaling laws on other systems is currently under way.

\acknowledgement
Partial funding for this research was provided by NSF through CAREER award DMR-1052808.\\

\bibliography{Critical}

\providecommand*\mcitethebibliography{\thebibliography}
\csname @ifundefined\endcsname{endmcitethebibliography}
  {\let\endmcitethebibliography\endthebibliography}{}
\begin{mcitethebibliography}{85}
\providecommand*\natexlab[1]{#1}
\providecommand*\mciteSetBstSublistMode[1]{}
\providecommand*\mciteSetBstMaxWidthForm[2]{}
\providecommand*\mciteBstWouldAddEndPuncttrue
  {\def\EndOfBibitem{\unskip.}}
\providecommand*\mciteBstWouldAddEndPunctfalse
  {\let\EndOfBibitem\relax}
\providecommand*\mciteSetBstMidEndSepPunct[3]{}
\providecommand*\mciteSetBstSublistLabelBeginEnd[3]{}
\providecommand*\EndOfBibitem{}
\mciteSetBstSublistMode{f}
\mciteSetBstMaxWidthForm{subitem}{(\alph{mcitesubitemcount})}
\mciteSetBstSublistLabelBeginEnd
  {\mcitemaxwidthsubitemform\space}
  {\relax}
  {\relax}

\bibitem[Bhatt et~al.(2006)Bhatt, Jasper, Schultz, Siepmann, and
  Truhlar]{bhatt2006critical}
Bhatt,~D.; Jasper,~A.~W.; Schultz,~N.~E.; Siepmann,~J.~I.; Truhlar,~D.~G.
  Critical properties of aluminum. \emph{J. Am. Chem. Soc.} \textbf{2006},
  \emph{128}, 4224--4225\relax
\mciteBstWouldAddEndPuncttrue
\mciteSetBstMidEndSepPunct{\mcitedefaultmidpunct}
{\mcitedefaultendpunct}{\mcitedefaultseppunct}\relax
\EndOfBibitem
\bibitem[Rowlinson and Swinton(1982)Rowlinson, and Swinton]{Rowlinson}
Rowlinson,~J.~S.; Swinton,~F.~L. \emph{Liquids and Liquid Mixtures};
  Butterworths, London, 1982\relax
\mciteBstWouldAddEndPuncttrue
\mciteSetBstMidEndSepPunct{\mcitedefaultmidpunct}
{\mcitedefaultendpunct}{\mcitedefaultseppunct}\relax
\EndOfBibitem
\bibitem[Delhommelle et~al.(1999)Delhommelle, Boutin, Tavitian, Mackie, and
  Fuchs]{FitCrit}
Delhommelle,~J.; Boutin,~A.; Tavitian,~B.; Mackie,~A.~D.; Fuchs,~A.~H.
  Vapour-liquid coexistence curves of the united-atom and anisotropic
  united-atom force fields for alkane mixtures. \emph{Mol. Phys.}
  \textbf{1999}, \emph{96}, 1517--1524\relax
\mciteBstWouldAddEndPuncttrue
\mciteSetBstMidEndSepPunct{\mcitedefaultmidpunct}
{\mcitedefaultendpunct}{\mcitedefaultseppunct}\relax
\EndOfBibitem
\bibitem[Potoff et~al.(1999)Potoff, Errington, and Panagiotopoulos]{Potoff}
Potoff,~J.; Errington,~J.; Panagiotopoulos,~A. Molecular simulation of phase
  equilibria for mixtures of polar and non-polar components. \emph{Mol. Phys.}
  \textbf{1999}, \emph{97}, 1073--1083\relax
\mciteBstWouldAddEndPuncttrue
\mciteSetBstMidEndSepPunct{\mcitedefaultmidpunct}
{\mcitedefaultendpunct}{\mcitedefaultseppunct}\relax
\EndOfBibitem
\bibitem[Errington et~al.(1998)Errington, Boulougouris, Economou,
  Panagiotopoulos, and Theodorou]{Errington}
Errington,~J.~R.; Boulougouris,~G.~C.; Economou,~I.~G.; Panagiotopoulos,~A.~Z.;
  Theodorou,~D.~N. Molecular simulation of phase equilibria for water-methane
  and water-ethane mixtures. \emph{J. Phys. Chem. B} \textbf{1998}, \emph{102},
  8865--8873\relax
\mciteBstWouldAddEndPuncttrue
\mciteSetBstMidEndSepPunct{\mcitedefaultmidpunct}
{\mcitedefaultendpunct}{\mcitedefaultseppunct}\relax
\EndOfBibitem
\bibitem[Messerly et~al.(2015)Messerly, Rowley, Knotts~IV, and
  Wilding]{messerly2015improved}
Messerly,~R.~A.; Rowley,~R.~L.; Knotts~IV,~T.~A.; Wilding,~W.~V. An improved
  statistical analysis for predicting the critical temperature and critical
  density with Gibbs ensemble Monte Carlo simulation. \emph{J. Chem. Phys.}
  \textbf{2015}, \emph{143}, 104101\relax
\mciteBstWouldAddEndPuncttrue
\mciteSetBstMidEndSepPunct{\mcitedefaultmidpunct}
{\mcitedefaultendpunct}{\mcitedefaultseppunct}\relax
\EndOfBibitem
\bibitem[Schnabel et~al.(2007)Schnabel, Vrabec, and Hasse]{Vrabec}
Schnabel,~T.; Vrabec,~J.; Hasse,~H. Unlike Lennard--Jones parameters for
  vapor--liquid equilibria. \emph{J. Mol. Liq.} \textbf{2007}, \emph{135},
  170--178\relax
\mciteBstWouldAddEndPuncttrue
\mciteSetBstMidEndSepPunct{\mcitedefaultmidpunct}
{\mcitedefaultendpunct}{\mcitedefaultseppunct}\relax
\EndOfBibitem
\bibitem[Panagiotopoulos(1987)]{panagiotopoulos1987direct}
Panagiotopoulos,~A.~Z. Direct determination of phase coexistence properties of
  fluids by Monte Carlo simulation in a new ensemble. \emph{Mol. Phys.}
  \textbf{1987}, \emph{61}, 813--826\relax
\mciteBstWouldAddEndPuncttrue
\mciteSetBstMidEndSepPunct{\mcitedefaultmidpunct}
{\mcitedefaultendpunct}{\mcitedefaultseppunct}\relax
\EndOfBibitem
\bibitem[Cailletet and Mathias(1886)Cailletet, and Mathias]{Cailletet}
Cailletet,~L.; Mathias,~E. \emph{C.R Acad. Sci.} \textbf{1886}, \emph{102},
  1202\relax
\mciteBstWouldAddEndPuncttrue
\mciteSetBstMidEndSepPunct{\mcitedefaultmidpunct}
{\mcitedefaultendpunct}{\mcitedefaultseppunct}\relax
\EndOfBibitem
\bibitem[Widom and Rowlinson(1970)Widom, and Rowlinson]{widom1970new}
Widom,~B.; Rowlinson,~J.~S. New model for the study of liquid--vapor phase
  transitions. \emph{J. Chem. Phys.} \textbf{1970}, \emph{52}, 1670--1684\relax
\mciteBstWouldAddEndPuncttrue
\mciteSetBstMidEndSepPunct{\mcitedefaultmidpunct}
{\mcitedefaultendpunct}{\mcitedefaultseppunct}\relax
\EndOfBibitem
\bibitem[N\"arger and Balzarini(1990)N\"arger, and Balzarini]{PhysRevB.42.6651}
N\"arger,~U.; Balzarini,~D.~A. Coexistence-curve diameter and critical density
  of xenon. \emph{Phys. Rev. B} \textbf{1990}, \emph{42}, 6651--6657\relax
\mciteBstWouldAddEndPuncttrue
\mciteSetBstMidEndSepPunct{\mcitedefaultmidpunct}
{\mcitedefaultendpunct}{\mcitedefaultseppunct}\relax
\EndOfBibitem
\bibitem[Weiner et~al.(1974)Weiner, Langley, and Ford]{PhysRevLett.32.879}
Weiner,~J.; Langley,~K.~H.; Ford,~N.~C. Experimental Evidence for a Departure
  from the Law of the Rectilinear Diameter. \emph{Phys. Rev. Lett.}
  \textbf{1974}, \emph{32}, 879--881\relax
\mciteBstWouldAddEndPuncttrue
\mciteSetBstMidEndSepPunct{\mcitedefaultmidpunct}
{\mcitedefaultendpunct}{\mcitedefaultseppunct}\relax
\EndOfBibitem
\bibitem[Fisher and Orkoulas(2000)Fisher, and Orkoulas]{PhysRevLett.85.696}
Fisher,~M.~E.; Orkoulas,~G. The Yang-Yang Anomaly in Fluid Criticality:
  Experiment and Scaling Theory. \emph{Phys. Rev. Lett.} \textbf{2000},
  \emph{85}, 696--699\relax
\mciteBstWouldAddEndPuncttrue
\mciteSetBstMidEndSepPunct{\mcitedefaultmidpunct}
{\mcitedefaultendpunct}{\mcitedefaultseppunct}\relax
\EndOfBibitem
\bibitem[Sengers and Shanks(2009)Sengers, and Shanks]{sengers2009experimental}
Sengers,~J.~V.; Shanks,~J.~G. Experimental critical-exponent values for fluids.
  \emph{J. Stat. Phys.} \textbf{2009}, \emph{137}, 857--877\relax
\mciteBstWouldAddEndPuncttrue
\mciteSetBstMidEndSepPunct{\mcitedefaultmidpunct}
{\mcitedefaultendpunct}{\mcitedefaultseppunct}\relax
\EndOfBibitem
\bibitem[J{\"u}ngst et~al.(1985)J{\"u}ngst, Knuth, and Hensel]{Jungst}
J{\"u}ngst,~S.; Knuth,~B.; Hensel,~F. Observation of singular diameters in the
  coexistence curves of metals. \emph{Phys. Rev. Lett.} \textbf{1985},
  \emph{55}, 2160\relax
\mciteBstWouldAddEndPuncttrue
\mciteSetBstMidEndSepPunct{\mcitedefaultmidpunct}
{\mcitedefaultendpunct}{\mcitedefaultseppunct}\relax
\EndOfBibitem
\bibitem[Goldstein and Ashcroft(1985)Goldstein, and Ashcroft]{Goldstein}
Goldstein,~R.~E.; Ashcroft,~N.~W. Origin of the Singular Diameter in the
  Coexistence Curve of a Metal. \emph{Phys. Rev. Lett.} \textbf{1985},
  \emph{55}, 2164\relax
\mciteBstWouldAddEndPuncttrue
\mciteSetBstMidEndSepPunct{\mcitedefaultmidpunct}
{\mcitedefaultendpunct}{\mcitedefaultseppunct}\relax
\EndOfBibitem
\bibitem[Schr{\"o}er and Pottlacher(2014)Schr{\"o}er, and
  Pottlacher]{schroer2014estimation}
Schr{\"o}er,~W.; Pottlacher,~G. Estimation of critical data and phase diagrams
  of pure molten metals. \emph{High Temperatures--High Pressures}
  \textbf{2014}, \emph{43}\relax
\mciteBstWouldAddEndPuncttrue
\mciteSetBstMidEndSepPunct{\mcitedefaultmidpunct}
{\mcitedefaultendpunct}{\mcitedefaultseppunct}\relax
\EndOfBibitem
\bibitem[Goldstein et~al.(1987)Goldstein, Parola, Ashcroft, Pestak, Chan,
  de~Bruyn, and Balzarini]{Goldstein2}
Goldstein,~R.~E.; Parola,~A.; Ashcroft,~N.; Pestak,~M.; Chan,~M.;
  de~Bruyn,~J.~R.; Balzarini,~D. Beyond the pair-potential model of fluids at
  the liquid-vapor critical point. \emph{Phys. Rev. Lett.} \textbf{1987},
  \emph{58}, 41\relax
\mciteBstWouldAddEndPuncttrue
\mciteSetBstMidEndSepPunct{\mcitedefaultmidpunct}
{\mcitedefaultendpunct}{\mcitedefaultseppunct}\relax
\EndOfBibitem
\bibitem[Bhatt et~al.(2006)Bhatt, Schultz, Jasper, Siepmann, and
  Truhlar]{bhatt2006phase}
Bhatt,~D.; Schultz,~N.~E.; Jasper,~A.~W.; Siepmann,~J.~I.; Truhlar,~D.~G. Phase
  behavior of elemental aluminum using Monte Carlo simulations. \emph{J. Phys.
  Chem. B} \textbf{2006}, \emph{110}, 26135--26142\relax
\mciteBstWouldAddEndPuncttrue
\mciteSetBstMidEndSepPunct{\mcitedefaultmidpunct}
{\mcitedefaultendpunct}{\mcitedefaultseppunct}\relax
\EndOfBibitem
\bibitem[Pottlacher and J{\"a}ger(1996)Pottlacher, and
  J{\"a}ger]{pottlacher1996review}
Pottlacher,~G.; J{\"a}ger,~H. A review of determinations of critical point data
  of metals using subsecond pulse heating techniques. \emph{J. Non-Cryst.
  Solids} \textbf{1996}, \emph{205}, 265--269\relax
\mciteBstWouldAddEndPuncttrue
\mciteSetBstMidEndSepPunct{\mcitedefaultmidpunct}
{\mcitedefaultendpunct}{\mcitedefaultseppunct}\relax
\EndOfBibitem
\bibitem[Pottlacher et~al.(1993)Pottlacher, Kaschnitz, and
  J{\"a}ger]{pottlacher1993investigations}
Pottlacher,~G.; Kaschnitz,~E.; J{\"a}ger,~H. Investigations of thermophysical
  properties of liquid metals with a rapid resistive heating technique.
  \emph{J. Non-Cryst. Solids} \textbf{1993}, \emph{156}, 374--378\relax
\mciteBstWouldAddEndPuncttrue
\mciteSetBstMidEndSepPunct{\mcitedefaultmidpunct}
{\mcitedefaultendpunct}{\mcitedefaultseppunct}\relax
\EndOfBibitem
\bibitem[Lang(2012)]{lang}
Lang,~G. \emph{IMetallkunde} \textbf{2012}, \emph{68}, 213\relax
\mciteBstWouldAddEndPuncttrue
\mciteSetBstMidEndSepPunct{\mcitedefaultmidpunct}
{\mcitedefaultendpunct}{\mcitedefaultseppunct}\relax
\EndOfBibitem
\bibitem[Chhabra(1990)]{chhabra1990surface}
Chhabra,~R. Surface tension of liquid metals: a predictive approach. \emph{High
  Temp. High Press.} \textbf{1990}, \emph{22}, 171--175\relax
\mciteBstWouldAddEndPuncttrue
\mciteSetBstMidEndSepPunct{\mcitedefaultmidpunct}
{\mcitedefaultendpunct}{\mcitedefaultseppunct}\relax
\EndOfBibitem
\bibitem[Kaptay(2012)]{kaptay}
Kaptay,~G. On the Order--Disorder Surface Phase Transition and Critical
  Temperature of Pure Metals Originating from BCC, FCC, and HCP Crystal
  Structures. \emph{Int. J. Thermophys.} \textbf{2012}, \emph{33},
  1177--1190\relax
\mciteBstWouldAddEndPuncttrue
\mciteSetBstMidEndSepPunct{\mcitedefaultmidpunct}
{\mcitedefaultendpunct}{\mcitedefaultseppunct}\relax
\EndOfBibitem
\bibitem[Morel et~al.(2009)Morel, Bultel, and Ch{\'e}ron]{morel2009critical}
Morel,~V.; Bultel,~A.; Ch{\'e}ron,~B. The critical temperature of aluminum.
  \emph{Int. J. Thermophys.} \textbf{2009}, \emph{30}, 1853--1863\relax
\mciteBstWouldAddEndPuncttrue
\mciteSetBstMidEndSepPunct{\mcitedefaultmidpunct}
{\mcitedefaultendpunct}{\mcitedefaultseppunct}\relax
\EndOfBibitem
\bibitem[Ramana(2014)]{ramana2014molecular}
Ramana,~A. S.~V. Molecular dynamics simulation of liquid--vapor phase diagrams
  of metals modeled using modified empirical pair potentials. \emph{Fluid Phase
  Equil.} \textbf{2014}, \emph{361}, 181--187\relax
\mciteBstWouldAddEndPuncttrue
\mciteSetBstMidEndSepPunct{\mcitedefaultmidpunct}
{\mcitedefaultendpunct}{\mcitedefaultseppunct}\relax
\EndOfBibitem
\bibitem[Singh et~al.(2006)Singh, Adhikari, and Kwak]{singh2006vapor}
Singh,~J.~K.; Adhikari,~J.; Kwak,~S.~K. Vapor--liquid phase coexistence curves
  for Morse fluids. \emph{Fluid Phase Equil.} \textbf{2006}, \emph{248},
  1--6\relax
\mciteBstWouldAddEndPuncttrue
\mciteSetBstMidEndSepPunct{\mcitedefaultmidpunct}
{\mcitedefaultendpunct}{\mcitedefaultseppunct}\relax
\EndOfBibitem
\bibitem[Apfelbaum and Vorob'ev(2015)Apfelbaum, and
  Vorob'ev]{apfelbaum2015wide}
Apfelbaum,~E.~M.; Vorob'ev,~V.~S. The Wide-Range Method to Construct the Entire
  Coexistence Liquid--Gas Curve and to Determine the Critical Parameters of
  Metals. \emph{J. Phys. Chem. B} \textbf{2015}, \emph{119}, 11825--11832\relax
\mciteBstWouldAddEndPuncttrue
\mciteSetBstMidEndSepPunct{\mcitedefaultmidpunct}
{\mcitedefaultendpunct}{\mcitedefaultseppunct}\relax
\EndOfBibitem
\bibitem[Apfelbaum(2012)]{apfelbaum2012estimate}
Apfelbaum,~E. Estimate of Beryllium Critical Point on the Basis of
  Correspondence between the Critical and the Zeno-Line Parameters. \emph{J.
  Phys. Chem. B} \textbf{2012}, \emph{116}, 14660--14666\relax
\mciteBstWouldAddEndPuncttrue
\mciteSetBstMidEndSepPunct{\mcitedefaultmidpunct}
{\mcitedefaultendpunct}{\mcitedefaultseppunct}\relax
\EndOfBibitem
\bibitem[Apfelbaum and Vorob'ev(2014)Apfelbaum, and
  Vorob'ev]{apfelbaum2014saturation}
Apfelbaum,~E.; Vorob'ev,~V. The saturation pressure for different objects in
  reduced variables and the justification of some empirical relations set from
  the van der Waals equation. \emph{Chem. Phys. Lett.} \textbf{2014},
  \emph{591}, 212--215\relax
\mciteBstWouldAddEndPuncttrue
\mciteSetBstMidEndSepPunct{\mcitedefaultmidpunct}
{\mcitedefaultendpunct}{\mcitedefaultseppunct}\relax
\EndOfBibitem
\bibitem[Apfelbaum and Vorob'ev(2008)Apfelbaum, and Vorob'ev]{apfelbaum2008new}
Apfelbaum,~E.; Vorob'ev,~V. A new similarity found from the correspondence of
  the critical and Zeno-line parameters. \emph{J. Phys. Chem. B} \textbf{2008},
  \emph{112}, 13064--13069\relax
\mciteBstWouldAddEndPuncttrue
\mciteSetBstMidEndSepPunct{\mcitedefaultmidpunct}
{\mcitedefaultendpunct}{\mcitedefaultseppunct}\relax
\EndOfBibitem
\bibitem[Apfelbaum et~al.(2006)Apfelbaum, Vorob'ev, and
  Martynov]{apfelbaum2006triangle}
Apfelbaum,~E.; Vorob'ev,~V.; Martynov,~G. Triangle of liquid-gas states.
  \emph{J. Phys. Chem. B} \textbf{2006}, \emph{110}, 8474--8480\relax
\mciteBstWouldAddEndPuncttrue
\mciteSetBstMidEndSepPunct{\mcitedefaultmidpunct}
{\mcitedefaultendpunct}{\mcitedefaultseppunct}\relax
\EndOfBibitem
\bibitem[Kulinskii(2010)]{kulinskii2010simple}
Kulinskii,~V. Simple geometrical interpretation of the linear character for the
  Zeno-line and the rectilinear diameter. \emph{J. Phys. Chem. B}
  \textbf{2010}, \emph{114}, 2852--2855\relax
\mciteBstWouldAddEndPuncttrue
\mciteSetBstMidEndSepPunct{\mcitedefaultmidpunct}
{\mcitedefaultendpunct}{\mcitedefaultseppunct}\relax
\EndOfBibitem
\bibitem[Kulinskii(2014)]{kulinskii2014critical}
Kulinskii,~V. The critical compressibility factor value: Associative fluids and
  liquid alkali metals. \emph{J. Chem. Phys.} \textbf{2014}, \emph{141},
  054503\relax
\mciteBstWouldAddEndPuncttrue
\mciteSetBstMidEndSepPunct{\mcitedefaultmidpunct}
{\mcitedefaultendpunct}{\mcitedefaultseppunct}\relax
\EndOfBibitem
\bibitem[Vorob'ev(2014)]{Vorobev}
Vorob'ev,~V. How to turn real substance liquid-gas coexistence curve in binodal
  of lattice gas. \emph{Chem. Phys. Lett,} \textbf{2014}, \emph{605},
  47--50\relax
\mciteBstWouldAddEndPuncttrue
\mciteSetBstMidEndSepPunct{\mcitedefaultmidpunct}
{\mcitedefaultendpunct}{\mcitedefaultseppunct}\relax
\EndOfBibitem
\bibitem[Luo et~al.(2003)Luo, Ahrens, {\c{C}}a{\u{g}}{\i}n, Strachan,
  Goddard~III, and Swift]{luo2003maximum}
Luo,~S.-N.; Ahrens,~T.~J.; {\c{C}}a{\u{g}}{\i}n,~T.; Strachan,~A.;
  Goddard~III,~W.~A.; Swift,~D.~C. Maximum superheating and undercooling:
  Systematics, molecular dynamics simulations, and dynamic experiments.
  \emph{Phys. Rev. B} \textbf{2003}, \emph{68}, 134206\relax
\mciteBstWouldAddEndPuncttrue
\mciteSetBstMidEndSepPunct{\mcitedefaultmidpunct}
{\mcitedefaultendpunct}{\mcitedefaultseppunct}\relax
\EndOfBibitem
\bibitem[Kart et~al.(2005)Kart, Tomak, Uludo{\u{g}}an, and
  {\c{C}}a{\u{g}}{\i}n]{kart2005thermodynamical}
Kart,~H.; Tomak,~M.; Uludo{\u{g}}an,~M.; {\c{C}}a{\u{g}}{\i}n,~T.
  Thermodynamical and mechanical properties of Pd--Ag alloys. \emph{Comput.
  Mat. Sci.} \textbf{2005}, \emph{32}, 107--117\relax
\mciteBstWouldAddEndPuncttrue
\mciteSetBstMidEndSepPunct{\mcitedefaultmidpunct}
{\mcitedefaultendpunct}{\mcitedefaultseppunct}\relax
\EndOfBibitem
\bibitem[Xu et~al.(2005)Xu, Cagin, and Goddard~III]{xu2005assessment}
Xu,~P.; Cagin,~T.; Goddard~III,~W.~A. Assessment of phenomenological models for
  viscosity of liquids based on nonequilibrium atomistic simulations of copper.
  \emph{J. Chem. Phys.} \textbf{2005}, \emph{123}, 104506\relax
\mciteBstWouldAddEndPuncttrue
\mciteSetBstMidEndSepPunct{\mcitedefaultmidpunct}
{\mcitedefaultendpunct}{\mcitedefaultseppunct}\relax
\EndOfBibitem
\bibitem[Qi et~al.(2001)Qi, {\c{C}}a{\u{g}}in, Kimura, and
  Goddard~Iii]{qi2001viscosities}
Qi,~Y.; {\c{C}}a{\u{g}}in,~T.; Kimura,~Y.; Goddard~Iii,~W.~A. Viscosities of
  liquid metal alloys from nonequilibrium molecular dynamics. \emph{Journal of
  computer-aided materials design} \textbf{2001}, \emph{8}, 233--243\relax
\mciteBstWouldAddEndPuncttrue
\mciteSetBstMidEndSepPunct{\mcitedefaultmidpunct}
{\mcitedefaultendpunct}{\mcitedefaultseppunct}\relax
\EndOfBibitem
\bibitem[Kart et~al.(2004)Kart, Tomak, Uludo{\u{g}}an, and
  {\c{C}}a{\u{g}}{\i}n]{kart2004liquid}
Kart,~S.~{\"O}.; Tomak,~M.; Uludo{\u{g}}an,~M.; {\c{C}}a{\u{g}}{\i}n,~T. Liquid
  properties of Pd--Ni alloys. \emph{J. Non-Cryst. Solids} \textbf{2004},
  \emph{337}, 101--108\relax
\mciteBstWouldAddEndPuncttrue
\mciteSetBstMidEndSepPunct{\mcitedefaultmidpunct}
{\mcitedefaultendpunct}{\mcitedefaultseppunct}\relax
\EndOfBibitem
\bibitem[Kart et~al.(2006)Kart, Tomak, Uludo{\u{g}}an, and
  {\c{C}}a{\u{g}}{\i}n]{kart2006structural}
Kart,~S.~{\"O}.; Tomak,~M.; Uludo{\u{g}}an,~M.; {\c{C}}a{\u{g}}{\i}n,~T.
  Structural, thermodynamical, and transport properties of undercooled binary
  Pd--Ni alloys. \emph{Mat. Sci. Eng. A} \textbf{2006}, \emph{435},
  736--744\relax
\mciteBstWouldAddEndPuncttrue
\mciteSetBstMidEndSepPunct{\mcitedefaultmidpunct}
{\mcitedefaultendpunct}{\mcitedefaultseppunct}\relax
\EndOfBibitem
\bibitem[Desgranges and Delhommelle(2007)Desgranges, and
  Delhommelle]{desgranges2007viscosity}
Desgranges,~C.; Delhommelle,~J. Viscosity of liquid iron under high pressure
  and high temperature: Equilibrium and nonequilibrium molecular dynamics
  simulation studies. \emph{Phys. Rev. B} \textbf{2007}, \emph{76},
  172102\relax
\mciteBstWouldAddEndPuncttrue
\mciteSetBstMidEndSepPunct{\mcitedefaultmidpunct}
{\mcitedefaultendpunct}{\mcitedefaultseppunct}\relax
\EndOfBibitem
\bibitem[Desgranges and Delhommelle(2008)Desgranges, and
  Delhommelle]{desgranges2008shear}
Desgranges,~C.; Delhommelle,~J. Shear viscosity of liquid copper at
  experimentally accessible shear rates: Application of the transient-time
  correlation function formalism. \emph{J. Chem. Phys.} \textbf{2008},
  \emph{128}, 084506\relax
\mciteBstWouldAddEndPuncttrue
\mciteSetBstMidEndSepPunct{\mcitedefaultmidpunct}
{\mcitedefaultendpunct}{\mcitedefaultseppunct}\relax
\EndOfBibitem
\bibitem[Desgranges and Delhommelle(2008)Desgranges, and
  Delhommelle]{desgranges2008rheology}
Desgranges,~C.; Delhommelle,~J. Rheology of liquid fcc metals: Equilibrium and
  transient-time correlation-function nonequilibrium molecular dynamics
  simulations. \emph{Phys. Rev. B} \textbf{2008}, \emph{78}, 184202\relax
\mciteBstWouldAddEndPuncttrue
\mciteSetBstMidEndSepPunct{\mcitedefaultmidpunct}
{\mcitedefaultendpunct}{\mcitedefaultseppunct}\relax
\EndOfBibitem
\bibitem[Gelb and Chakraborty(2011)Gelb, and Chakraborty]{gelb2011boiling}
Gelb,~L.~D.; Chakraborty,~S.~N. Boiling point determination using adiabatic
  Gibbs ensemble Monte Carlo simulations: Application to metals described by
  embedded-atom potentials. \emph{J. Chem. Phys.} \textbf{2011}, \emph{135},
  224113\relax
\mciteBstWouldAddEndPuncttrue
\mciteSetBstMidEndSepPunct{\mcitedefaultmidpunct}
{\mcitedefaultendpunct}{\mcitedefaultseppunct}\relax
\EndOfBibitem
\bibitem[Desgranges et~al.(2010)Desgranges, Kastl, Aleksandrov, and
  Delhommelle]{Tsvetan}
Desgranges,~C.; Kastl,~E.~A.; Aleksandrov,~T.; Delhommelle,~J. Optimisation of
  Multiple Time-Step Hybrid Monte Carlo Wang--Landau Simulations in the
  Isobaric--Isothermal Ensemble for the Determination of Phase Equilibria.
  \emph{Molec. Simul.} \textbf{2010}, \emph{36}, 544--551\relax
\mciteBstWouldAddEndPuncttrue
\mciteSetBstMidEndSepPunct{\mcitedefaultmidpunct}
{\mcitedefaultendpunct}{\mcitedefaultseppunct}\relax
\EndOfBibitem
\bibitem[Aleksandrov et~al.(2012)Aleksandrov, Desgranges, and
  Delhommelle]{Tsvetan2}
Aleksandrov,~T.; Desgranges,~C.; Delhommelle,~J. Numerical Estimate for Boiling
  Points via Wang-Landau Simulations. \emph{Molec. Simul.} \textbf{2012},
  \emph{38}, 1265--1270\relax
\mciteBstWouldAddEndPuncttrue
\mciteSetBstMidEndSepPunct{\mcitedefaultmidpunct}
{\mcitedefaultendpunct}{\mcitedefaultseppunct}\relax
\EndOfBibitem
\bibitem[Desgranges and Delhommelle(2012)Desgranges, and Delhommelle]{PartI}
Desgranges,~C.; Delhommelle,~J. Evaluation of the Grand-Canonical Partition
  Function using Expanded Wang-Landau Simulations. I. Thermodynamic Properties
  in the Bulk and at the Liquid-Vapor Phase Boundary. \emph{J. Chem. Phys.}
  \textbf{2012}, \emph{136}, 184107\relax
\mciteBstWouldAddEndPuncttrue
\mciteSetBstMidEndSepPunct{\mcitedefaultmidpunct}
{\mcitedefaultendpunct}{\mcitedefaultseppunct}\relax
\EndOfBibitem
\bibitem[Desgranges and Delhommelle(2012)Desgranges, and Delhommelle]{PartII}
Desgranges,~C.; Delhommelle,~J. Evaluation of the Grand-Canonical Partition
  Function using Expanded Wang-Landau Simulations. II. Adsorption of Atomic and
  Molecular Fluids in a Porous Material. \emph{J. Chem. Phys.} \textbf{2012},
  \emph{136}, 184108\relax
\mciteBstWouldAddEndPuncttrue
\mciteSetBstMidEndSepPunct{\mcitedefaultmidpunct}
{\mcitedefaultendpunct}{\mcitedefaultseppunct}\relax
\EndOfBibitem
\bibitem[Desgranges and Delhommelle(2012)Desgranges, and Delhommelle]{PartIII}
Desgranges,~C.; Delhommelle,~J. Evaluation of the Grand-Canonical Partition
  Function using Expanded Wang-Landau Simulations. III. Adsorption of Atomic
  and Molecular Fluids in a Porous Material. \emph{J. Chem. Phys.}
  \textbf{2012}, \emph{136}, 184108\relax
\mciteBstWouldAddEndPuncttrue
\mciteSetBstMidEndSepPunct{\mcitedefaultmidpunct}
{\mcitedefaultendpunct}{\mcitedefaultseppunct}\relax
\EndOfBibitem
\bibitem[Desgranges and Delhommelle(2016)Desgranges, and Delhommelle]{PartIV}
Desgranges,~C.; Delhommelle,~J. Evaluation of the grand-canonical partition
  function using expanded Wang-Landau simulations. IV. Performance of many-body
  force fields and tight-binding schemes for the fluid phases of silicon.
  \emph{J. Chem. Phys.} \textbf{2016}, \emph{144}, 124510\relax
\mciteBstWouldAddEndPuncttrue
\mciteSetBstMidEndSepPunct{\mcitedefaultmidpunct}
{\mcitedefaultendpunct}{\mcitedefaultseppunct}\relax
\EndOfBibitem
\bibitem[Wang and Landau(2001)Wang, and Landau]{Wang1}
Wang,~F.; Landau,~D.~P. Determining the Density of States for Classical
  Statistical Models: A Random Walk Algorithm to produce a Flat Histogram.
  \emph{Phys. Rev. E} \textbf{2001}, \emph{64}, 056101\relax
\mciteBstWouldAddEndPuncttrue
\mciteSetBstMidEndSepPunct{\mcitedefaultmidpunct}
{\mcitedefaultendpunct}{\mcitedefaultseppunct}\relax
\EndOfBibitem
\bibitem[Wang and Landau(2001)Wang, and Landau]{Wang2}
Wang,~F.; Landau,~D. Efficient, Multiple-Range Random Walk Algorithm to
  calculate the Density of States. \emph{Phys. Rev. Lett.} \textbf{2001},
  \emph{86}, 2050--2053\relax
\mciteBstWouldAddEndPuncttrue
\mciteSetBstMidEndSepPunct{\mcitedefaultmidpunct}
{\mcitedefaultendpunct}{\mcitedefaultseppunct}\relax
\EndOfBibitem
\bibitem[Shell et~al.(2002)Shell, Debenedetti, and Panagiotopoulos]{Shell}
Shell,~M.~S.; Debenedetti,~P.~G.; Panagiotopoulos,~A.~Z. Generalization of the
  Wang-Landau Method for Off-Lattice Simulations. \emph{Phys. Rev. E}
  \textbf{2002}, \emph{66}, 056703\relax
\mciteBstWouldAddEndPuncttrue
\mciteSetBstMidEndSepPunct{\mcitedefaultmidpunct}
{\mcitedefaultendpunct}{\mcitedefaultseppunct}\relax
\EndOfBibitem
\bibitem[Yan et~al.(2002)Yan, Faller, and de~Pablo]{Yan}
Yan,~Q.; Faller,~R.; de~Pablo,~J.~J. Density-of-States Monte Carlo Method for
  Simulation of Fluids. \emph{J. Chem. Phys.} \textbf{2002}, \emph{116},
  8745--8750\relax
\mciteBstWouldAddEndPuncttrue
\mciteSetBstMidEndSepPunct{\mcitedefaultmidpunct}
{\mcitedefaultendpunct}{\mcitedefaultseppunct}\relax
\EndOfBibitem
\bibitem[Gazenm$\ddot{\mathrm{u}}$ller and
  Camp(2007)Gazenm$\ddot{\mathrm{u}}$ller, and Camp]{Camp}
Gazenm$\ddot{\mathrm{u}}$ller,~G.; Camp,~P.~J. Applications of Wang-Landau
  Sampling to Determine Phase Equilibria in Complex Fluids. \emph{J. Chem.
  Phys.} \textbf{2007}, \emph{127}, 154504\relax
\mciteBstWouldAddEndPuncttrue
\mciteSetBstMidEndSepPunct{\mcitedefaultmidpunct}
{\mcitedefaultendpunct}{\mcitedefaultseppunct}\relax
\EndOfBibitem
\bibitem[Desgranges and Delhommelle(2009)Desgranges, and Delhommelle]{WLHMC}
Desgranges,~C.; Delhommelle,~J. Phase Equilibria of Molecular Fluids via Hybrid
  Monte Carlo Wang-Landau Simulations: Applications to Benzene and n-Alkanes.
  \emph{J. Chem. Phys.} \textbf{2009}, \emph{130}, 244109\relax
\mciteBstWouldAddEndPuncttrue
\mciteSetBstMidEndSepPunct{\mcitedefaultmidpunct}
{\mcitedefaultendpunct}{\mcitedefaultseppunct}\relax
\EndOfBibitem
\bibitem[Desgranges et~al.(2010)Desgranges, Hicks, Magness, and
  Delhommelle]{DHMD}
Desgranges,~C.; Hicks,~J.~M.; Magness,~A.; Delhommelle,~J. Phase Equilibria of
  Polyaromatic Hydrocarbons by Hybrid Monte Carlo Wang--Landau Simulations.
  \emph{Mol. Phys.} \textbf{2010}, \emph{108}, 151--158\relax
\mciteBstWouldAddEndPuncttrue
\mciteSetBstMidEndSepPunct{\mcitedefaultmidpunct}
{\mcitedefaultendpunct}{\mcitedefaultseppunct}\relax
\EndOfBibitem
\bibitem[Ngale et~al.(2012)Ngale, Desgranges, and Delhommelle]{KennethI}
Ngale,~K.~N.; Desgranges,~C.; Delhommelle,~J. Wang-Landau Configurational Bias
  Monte Carlo Simulations: Vapour-Liquid Equilibria of Alkenes. \emph{Molec.
  Simul.} \textbf{2012}, \emph{38}, 653--658\relax
\mciteBstWouldAddEndPuncttrue
\mciteSetBstMidEndSepPunct{\mcitedefaultmidpunct}
{\mcitedefaultendpunct}{\mcitedefaultseppunct}\relax
\EndOfBibitem
\bibitem[Desgranges et~al.(2012)Desgranges, Ngale, and Delhommelle]{KennethII}
Desgranges,~C.; Ngale,~K.; Delhommelle,~J. Prediction of Critical Properties
  for Naphthacene, Triphenylene and Chrysene by Wang-Landau Simulations.
  \emph{Fluid Phase Equil.} \textbf{2012}, \emph{322-323}, 92--96\relax
\mciteBstWouldAddEndPuncttrue
\mciteSetBstMidEndSepPunct{\mcitedefaultmidpunct}
{\mcitedefaultendpunct}{\mcitedefaultseppunct}\relax
\EndOfBibitem
\bibitem[Escobedo and de~Pablo(1996)Escobedo, and de~Pablo]{expanded}
Escobedo,~F.; de~Pablo,~J.~J. Expanded Grand Canonical and Gibbs Ensemble Monte
  Carlo Simulation of Polymers. \emph{J. Chem. Phys.} \textbf{1996},
  \emph{105}, 4391\relax
\mciteBstWouldAddEndPuncttrue
\mciteSetBstMidEndSepPunct{\mcitedefaultmidpunct}
{\mcitedefaultendpunct}{\mcitedefaultseppunct}\relax
\EndOfBibitem
\bibitem[Lyubartsev et~al.(1992)Lyubartsev, Martsinovski, Shevkunov, and
  Vorontsov-Velyaminov]{Lyubartsev}
Lyubartsev,~A.~P.; Martsinovski,~A.~A.; Shevkunov,~S.~V.;
  Vorontsov-Velyaminov,~P.~N. New Approach to Monte Carlo Calculation of the
  Free Energy: Method of Expanded Ensembles. \emph{J. Chem. Phys.}
  \textbf{1992}, \emph{96}, 1776--1783\relax
\mciteBstWouldAddEndPuncttrue
\mciteSetBstMidEndSepPunct{\mcitedefaultmidpunct}
{\mcitedefaultendpunct}{\mcitedefaultseppunct}\relax
\EndOfBibitem
\bibitem[Muller and Paul(1994)Muller, and Paul]{Paul}
Muller,~M.; Paul,~W. Measuring the Chemical Potential of Polymer Solutions and
  Melts in Computer Simulations. \emph{J. Chem. Phys.} \textbf{1994},
  \emph{100}, 719--724\relax
\mciteBstWouldAddEndPuncttrue
\mciteSetBstMidEndSepPunct{\mcitedefaultmidpunct}
{\mcitedefaultendpunct}{\mcitedefaultseppunct}\relax
\EndOfBibitem
\bibitem[Shi and Maginn(2007)Shi, and Maginn]{Shi}
Shi,~W.; Maginn,~E.~J. Continuous Fractional Component Monte Carlo: An Adaptive
  Biasing Method for Open System Atomistic Simulations. \emph{J. Chem. Theory
  Comp.} \textbf{2007}, \emph{3}, 1451--1463\relax
\mciteBstWouldAddEndPuncttrue
\mciteSetBstMidEndSepPunct{\mcitedefaultmidpunct}
{\mcitedefaultendpunct}{\mcitedefaultseppunct}\relax
\EndOfBibitem
\bibitem[Singh and Errington(2006)Singh, and Errington]{Singh}
Singh,~J.~K.; Errington,~J.~R. Calculation of Phase Coexistence Properties and
  Surface Tensions of n-Alkanes with Grand-Canonical Transition-Matrix Monte
  Carlo Simulation and Finite-Size Scaling. \emph{J. Phys. Chem. B}
  \textbf{2006}, \emph{110}, 1369--1376\relax
\mciteBstWouldAddEndPuncttrue
\mciteSetBstMidEndSepPunct{\mcitedefaultmidpunct}
{\mcitedefaultendpunct}{\mcitedefaultseppunct}\relax
\EndOfBibitem
\bibitem[Escobedo and Martinez-Veracoechea(2007)Escobedo, and
  Martinez-Veracoechea]{MV1}
Escobedo,~F.~A.; Martinez-Veracoechea,~F.~J. Optimized Expanded Ensembles for
  Simulations involving Molecular Insertions and Deletions. I. Closed systems.
  \emph{J. Chem. Phys.} \textbf{2007}, \emph{127}, 174103\relax
\mciteBstWouldAddEndPuncttrue
\mciteSetBstMidEndSepPunct{\mcitedefaultmidpunct}
{\mcitedefaultendpunct}{\mcitedefaultseppunct}\relax
\EndOfBibitem
\bibitem[Escobedo and Martinez-Veracoechea(2008)Escobedo, and
  Martinez-Veracoechea]{MV2}
Escobedo,~F.~A.; Martinez-Veracoechea,~F.~J. Optimization of Expanded Ensemble
  Methods. \emph{J. Chem. Phys.} \textbf{2008}, \emph{129}, 154107\relax
\mciteBstWouldAddEndPuncttrue
\mciteSetBstMidEndSepPunct{\mcitedefaultmidpunct}
{\mcitedefaultendpunct}{\mcitedefaultseppunct}\relax
\EndOfBibitem
\bibitem[Rane et~al.(2013)Rane, Murali, and Errington]{Rane1}
Rane,~K.~S.; Murali,~S.; Errington,~J.~R. Monte Carlo Simulation Methods for
  Computing Liquid--Vapor Saturation Properties of Model Systems. \emph{J.
  Chem. Theory Comput.} \textbf{2013}, \emph{9}, 2552--2566\relax
\mciteBstWouldAddEndPuncttrue
\mciteSetBstMidEndSepPunct{\mcitedefaultmidpunct}
{\mcitedefaultendpunct}{\mcitedefaultseppunct}\relax
\EndOfBibitem
\bibitem[Rane and Errington(2013)Rane, and Errington]{Rane2}
Rane,~K.~S.; Errington,~J.~R. Using Monte Carlo Simulation to Compute
  Liquid--Vapor Saturation Properties of Ionic Liquids. \emph{J. Phys. Chem. B}
  \textbf{2013}, \emph{117}, 8018--8030\relax
\mciteBstWouldAddEndPuncttrue
\mciteSetBstMidEndSepPunct{\mcitedefaultmidpunct}
{\mcitedefaultendpunct}{\mcitedefaultseppunct}\relax
\EndOfBibitem
\bibitem[Koenig et~al.(2014)Koenig, Desgranges, and Delhommelle]{Aaron}
Koenig,~A. R.~V.; Desgranges,~C.; Delhommelle,~J. Adsorption of Hydrogen in
  Covalent Organic Frameworks using Expanded Wang--Landau Simulations.
  \emph{Molec. Simul.} \textbf{2014}, \emph{40}, 71--79\relax
\mciteBstWouldAddEndPuncttrue
\mciteSetBstMidEndSepPunct{\mcitedefaultmidpunct}
{\mcitedefaultendpunct}{\mcitedefaultseppunct}\relax
\EndOfBibitem
\bibitem[Hicks et~al.(2014)Hicks, Desgranges, and Delhommelle]{Erica}
Hicks,~E.~A.; Desgranges,~C.; Delhommelle,~J. Adsorption and Diffusion of the
  Antiparkinsonian Drug Amantadine in Carbon Nanotubes. \emph{Molec. Simul.}
  \textbf{2014}, \emph{40}, 656--663\relax
\mciteBstWouldAddEndPuncttrue
\mciteSetBstMidEndSepPunct{\mcitedefaultmidpunct}
{\mcitedefaultendpunct}{\mcitedefaultseppunct}\relax
\EndOfBibitem
\bibitem[Owen et~al.(2015)Owen, Desgranges, and Delhommelle]{Andrew}
Owen,~A.~N.; Desgranges,~C.; Delhommelle,~J. A new force field for $H_2S$ and
  its binary and ternary mixtures with $CO_2$ and $CH_4$. \emph{Fluid Phase
  Equil.} \textbf{2015}, \emph{402}, 69--77\relax
\mciteBstWouldAddEndPuncttrue
\mciteSetBstMidEndSepPunct{\mcitedefaultmidpunct}
{\mcitedefaultendpunct}{\mcitedefaultseppunct}\relax
\EndOfBibitem
\bibitem[Desgranges and Delhommelle(2015)Desgranges, and Delhommelle]{jctc2015}
Desgranges,~C.; Delhommelle,~J. Many-body effects on the thermodynamics of
  fluids, mixtures and nanoconfined fluids. \emph{J. Chem. Theory Comput.}
  \textbf{2015}, \emph{11}, 5401\relax
\mciteBstWouldAddEndPuncttrue
\mciteSetBstMidEndSepPunct{\mcitedefaultmidpunct}
{\mcitedefaultendpunct}{\mcitedefaultseppunct}\relax
\EndOfBibitem
\bibitem[McQuarrie(1976)]{McQuarrie}
McQuarrie,~D.~A. \emph{Statistical Mechanics}; Harper \& Row, New York,
  1976\relax
\mciteBstWouldAddEndPuncttrue
\mciteSetBstMidEndSepPunct{\mcitedefaultmidpunct}
{\mcitedefaultendpunct}{\mcitedefaultseppunct}\relax
\EndOfBibitem
\bibitem[Finnis and Sinclair(1984)Finnis, and Sinclair]{finnis1984simple}
Finnis,~M.; Sinclair,~J. A simple empirical N-body potential for transition
  metals. \emph{Phil. Mag. A} \textbf{1984}, \emph{50}, 45--55\relax
\mciteBstWouldAddEndPuncttrue
\mciteSetBstMidEndSepPunct{\mcitedefaultmidpunct}
{\mcitedefaultendpunct}{\mcitedefaultseppunct}\relax
\EndOfBibitem
\bibitem[Sutton and Chen(1990)Sutton, and Chen]{sutton1990long}
Sutton,~A.; Chen,~J. Long-range finnis--sinclair potentials. \emph{Phil. Mag.
  Lett.} \textbf{1990}, \emph{61}, 139--146\relax
\mciteBstWouldAddEndPuncttrue
\mciteSetBstMidEndSepPunct{\mcitedefaultmidpunct}
{\mcitedefaultendpunct}{\mcitedefaultseppunct}\relax
\EndOfBibitem
\bibitem[Mei et~al.(1991)Mei, Davenport, and Fernando]{mei1991analytic}
Mei,~J.; Davenport,~J.; Fernando,~G. Analytic embedded-atom potentials for fcc
  metals: Application to liquid and solid copper. \emph{Phys. Rev. B}
  \textbf{1991}, \emph{43}, 4653\relax
\mciteBstWouldAddEndPuncttrue
\mciteSetBstMidEndSepPunct{\mcitedefaultmidpunct}
{\mcitedefaultendpunct}{\mcitedefaultseppunct}\relax
\EndOfBibitem
\bibitem[Daw and Baskes(1983)Daw, and Baskes]{daw1983semiempirical}
Daw,~M.~S.; Baskes,~M.~I. Semiempirical, quantum mechanical calculation of
  hydrogen embrittlement in metals. \emph{Phys. Rev. Lett.} \textbf{1983},
  \emph{50}, 1285\relax
\mciteBstWouldAddEndPuncttrue
\mciteSetBstMidEndSepPunct{\mcitedefaultmidpunct}
{\mcitedefaultendpunct}{\mcitedefaultseppunct}\relax
\EndOfBibitem
\bibitem[Fokin and Popov(2013)Fokin, and Popov]{fokin2013general}
Fokin,~L.~R.; Popov,~V.~N. General function of the unit compressibility factor
  for liquid and gaseous mercury. \emph{High Temperature} \textbf{2013},
  \emph{51}, 465--468\relax
\mciteBstWouldAddEndPuncttrue
\mciteSetBstMidEndSepPunct{\mcitedefaultmidpunct}
{\mcitedefaultendpunct}{\mcitedefaultseppunct}\relax
\EndOfBibitem
\bibitem[Apfelbaum and Vorob'ev(2009)Apfelbaum, and
  Vorob'ev]{apfelbaum2009predictions}
Apfelbaum,~E.; Vorob'ev,~V. The predictions of the critical point parameters
  for Al, Cu and W found from the correspondence between the critical point and
  unit compressibility line (Zeno line) positions. \emph{Chem. Phys. Lett.}
  \textbf{2009}, \emph{467}, 318--322\relax
\mciteBstWouldAddEndPuncttrue
\mciteSetBstMidEndSepPunct{\mcitedefaultmidpunct}
{\mcitedefaultendpunct}{\mcitedefaultseppunct}\relax
\EndOfBibitem
\bibitem[Geiger et~al.(1987)Geiger, Busse, and Loehrke]{geiger1987vapor}
Geiger,~F.; Busse,~C.; Loehrke,~R. The vapor pressure of indium, silver,
  gallium, copper, tin, and gold between 0.1 and 3.0 bar. \emph{Int. J.
  Thermophys.} \textbf{1987}, \emph{8}, 425--436\relax
\mciteBstWouldAddEndPuncttrue
\mciteSetBstMidEndSepPunct{\mcitedefaultmidpunct}
{\mcitedefaultendpunct}{\mcitedefaultseppunct}\relax
\EndOfBibitem
\bibitem[Linde(1994)]{linde1994handbook}
Linde,~D. Handbook of Chemistry and Physics. CRC Press. \emph{Boca Raton}
  \textbf{1994}, \relax
\mciteBstWouldAddEndPunctfalse
\mciteSetBstMidEndSepPunct{\mcitedefaultmidpunct}
{}{\mcitedefaultseppunct}\relax
\EndOfBibitem
\bibitem[Watson(1943)]{watson1943thermodynamics}
Watson,~K. Thermodynamics of the liquid state. \emph{Ind. Eng. Chem.}
  \textbf{1943}, \emph{35}, 398--406\relax
\mciteBstWouldAddEndPuncttrue
\mciteSetBstMidEndSepPunct{\mcitedefaultmidpunct}
{\mcitedefaultendpunct}{\mcitedefaultseppunct}\relax
\EndOfBibitem
\bibitem[Leibovici and Nichita(2014)Leibovici, and Nichita]{leibovici2014new}
Leibovici,~C.~F.; Nichita,~D.~V. New basis functions for the representation of
  vapor pressure data. \emph{Fluid Phase Equil.} \textbf{2014}, \emph{361},
  1--15\relax
\mciteBstWouldAddEndPuncttrue
\mciteSetBstMidEndSepPunct{\mcitedefaultmidpunct}
{\mcitedefaultendpunct}{\mcitedefaultseppunct}\relax
\EndOfBibitem
\end{mcitethebibliography}

\break

\section{Graphical Abstract}

\begin{scheme}
\includegraphics*[width=8cm]{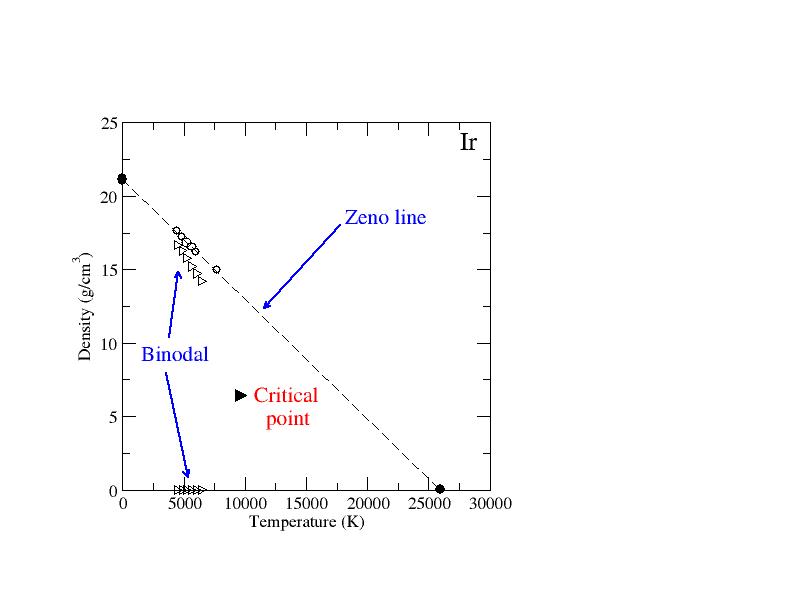}
\end{scheme}

\end{document}